\documentclass[12pt]{article}
\usepackage{graphicx}
\begin{document}
\centerline{ Cellular Automata Simulation of Medication-Induced Autoimmune Diseases}

\bigskip

Dietrich Stauffer and Ana Proykova$^1$ and 

Institute for Theoretical Physics, Cologne University, 

D-50923 K\"oln, Euroland

\bigskip
\noindent
$^1$ Department of Atomic Physics, University of Sofia,
5 James Bourchier Blvd., Sofia-1126, Bulgaria

\bigskip

Abstract:
We implement the cellular automata model proposed by Stauffer and Weisbuch
in 1992 to describe the response of the immune system to antigens in the
presence of medications. The model contains two thresholds, $\theta_1$
and $\theta_2$, suggested by de Boer, Segel, and Perelson to present the minimum
field needed to stimulate the proliferation of the receptors and to suppress
it, respectively. 
The influence of the drug is mimicked by
increasing the second threshold, thus enhancing the immune response. If this 
increase is too strong, the immune response is triggered in the whole 
immune repertoire, causing it to attack the own body. This effect is seen
in our simulations to depend both on the ratio of the thresholds
and on their absolute values.
 
Keywords: cellular automata, phase transitions,
thresholds, medical treatment, BSP model

\section{Introduction}

Too much of a good think may be bad. We know this from alcoholic drinks;
few readers have lost as much money as Bill Gates did in 2000; and antibiotics
not only saved many human lives but also created resistant strains of bacteria.
Our immune system protects us against many diseases but may also go wrong and 
attack the own body, leading e.g.  to
many allergic reactions. Allergic reactions are unexpected reactions that
 are not caused by the normal action of the medicine and are due to 
stimulation of the immune system by the drug.  The immune system
may react in a variety of ways, most commonly by causing proliferation
 of lymphocytes which form antibodies. Antibodies may cause reactions
 when they combine with the drug.

 The present note tries
to model something similar: If we apply too much of a medicine strengthening
our immune system, can it cause the immune cells to attack ``everything''
including healthy tissue, besides attacking the foreign antigen (infectious
disease)? The antigen-specific immunoglobulin interacts with 
mast cells to protect the host against the invading 
parasite. However, the same antibody-cell combination is also responsible 
for typical allergy or immediate hyper-sensitivity reactions such as hay 
fever, asthma, hives and anaphylaxis.

Presumably many immunological models \cite{perelson,zorzenonHIV,hershberg} 
would give such an effect.
To avoid adding another example to the already large number of immunology
models of the last decades, we selected a variant of the de Boer-Segel-Perelson
model \cite{bsp}, which has been cited well in the last decade. We use one of
the simplifications of \cite{weisbuch} called BSP II there; since an even 
simpler version, BSP III, has been criticised \cite{zorzenon}, we take a
somewhat more complicated and realistic version BSP II and explain it in the
next section. Thereafter we bring our results and summarise our findings.

\section{Model}
The immune repertoire is represented by a five-dimensional shape space of 
linear dimension $L$ containing $L^5$ possible types of antibodies, which the
immune cells may produce to fight against as many possible types of antigens. 
Each of the five dimensions corresponds to give different criteria (length,
electric charge, curvature, ...) by which our immune system characterises the
antigen; for L = 7 this parameter may take the values $x = -3, -2, -1, 0, 1, 
2, 3$. The five components $x$ then give a five-dimenensional vector ${\bf r}$
such that the null vector is the lattice centre. Just as a key has to fit a 
lock, apart from minor scratches, the antibody has to be complementary to the
antigen in order to detect and neutralise it. Thus an antigen with $x=-2$ 
may be killed by an antibody with $x=+2$; more generally an antigen at a
 position --{\bf r} in the shape space is neutralised by its
 complementary antibody at 
a position {\bf r}, or by very similar antibodies.
 We define as very similar the 
ten nearest neighbours of {\bf r}, which differ
from the fully complementary site {\bf r} only by a Hamming distance of one; 
but each of these ten neighbours
contributes only one tenth as much to the immune response as the fully 
complementary {\bf r}. Thus if $b({\bf r})$ is the concentration of B cells 
(producing antibodies) at a site ${\bf r}$ of the shape space, then 

$$h(-{\bf r}) = b({\bf r}) + 0.1 \sum_n b({\bf r_n}) \eqno (1) $$
is the influence of the immune system on the site --{\bf r};
 the sum runs over the
ten neighbours ${\bf r_n}$ of {\bf r}. 

Following the Weber-Fechner law of physiology, that effects increase often 
only logarithmic with concentration or physical amplitude, we assume the 
concentrations $b$ to vary exponentially:

$$b = e^B \eqno (2)$$
where the B's are integers varying between 0 and 20. The rule for our 
simultaneous updating is: $B$ increases by one if and only if the influence
$h$ at that site in the shape space lies between two thresholds $\theta_1$ and
$\theta_2$; otherwise it decreases by one. (If $B$ is zero it stays at zero;
parameters were chosen such that $B$ does not surpass 20). The ``naive''
immune system thus has $B = 0$ everywhere, i.e. only one B cell per a possible
shape. This threshold automata rule mimics the bell-shaped natural immune 
response. Now we completed the definition of  model BSP II. 

(Actually the BSP model does not deal with antigen-antibody reactions but
with the underlying idiotypic-antiidiotypic reactions within the immune 
network; therefore a field $h$ of the proper value $\theta_1 < h < \theta_2$
increases and not decreases the corresponding $b$ and $B$.)

\begin{figure}[hbt]
\begin{center}
\includegraphics[angle=-90,scale=0.5]{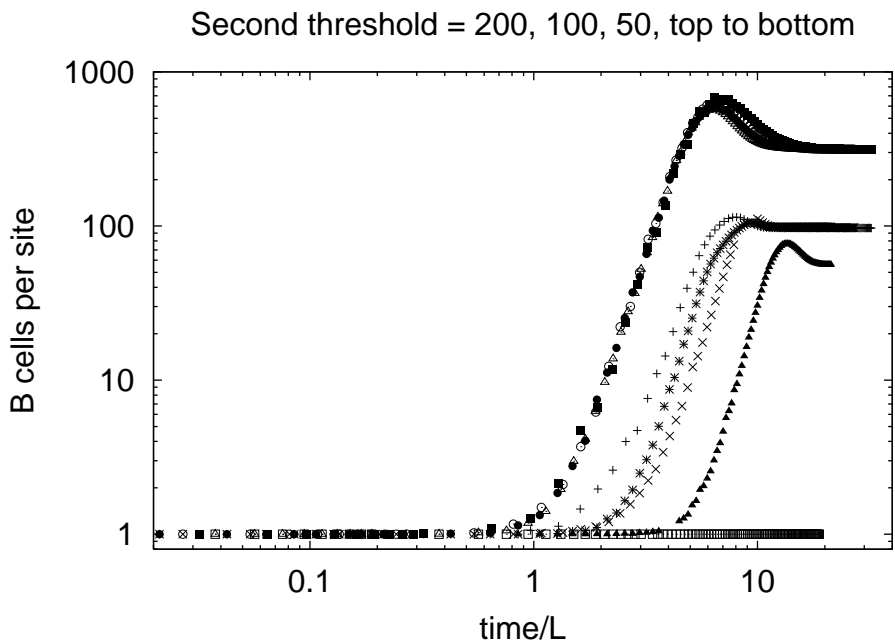}
\end{center}
\caption{Double-logarithmic plot of normalised response $B/L^5$ versus scaled 
time $t/L$ where $t$ is the number of sweeps through the lattice. Different
symbols correspond to different lattice sizes $L = 31$, 37, 47, and 53; 
$\theta_1=10$ throughout. The upper curves have $\theta_2=200$, the intermediate
curves $\theta_2  = 100$, and the flat line near unity mostly $\theta_2 = 
50$. However, one of the $\theta_2 = 100$ curves stays flat, and one of the
$\theta_2 = 50$ simulations rises and is the lowest curve in this figure. 
}
\end{figure}

\begin{figure}[hbt]
\begin{center}
\includegraphics[angle=-90,scale=0.5]{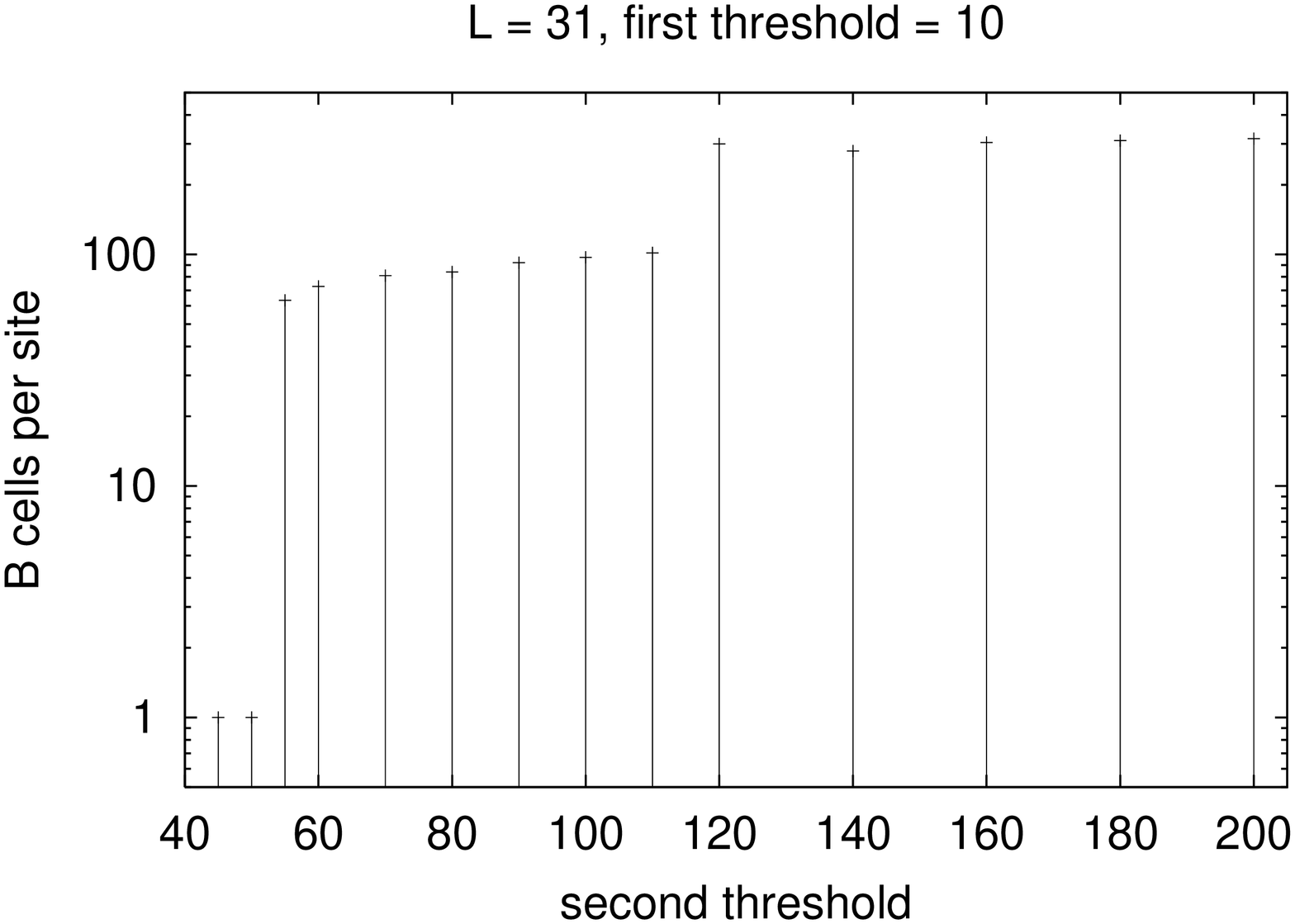}
\end{center}
\caption{Semilogarithmic plot of the number $S/L^5$ of B cells per site as
a function of $\theta_2$ for fixed $\theta_1 = 10$ and $L = 31$. We see first a
jump (first-order phase transition)
from unity to about 100, and then a further jump to about 300.
Only in the small left region the immune system works properly, with the first
jump it deteriorates, and with the second jump it becomes really bad.
}
\end{figure}

Initially we put the whole lattice to $B=0$ except at one line of length $L$ or 
one hyper-plane of $L^4$ sites; this initialisation models the initial response
to a specific antigen. We then ask whether this initial response remains small 
and localised around the initial region, or becomes large by spreading over
the whole lattice. We thus monitor the sum 

$$S = \sum_{\bf r} b({\bf r}) \eqno (3) $$
over the whole lattice; since our initially excited region is relatively small,
it does not matter whether or not we subtract from this overall sum the sum 
over only the initially excited region. Thus a healthy localised immune response
means $S/L^5 \simeq 1$ while a dangerous over-reaction means $S/L^5 \gg 1$. Our
simulations below show a clear separation between these two behaviours.
Fortran programs are available from anap@phys.uni-sofia.bg and 
stauffer@thp.uni-koeln.de.

To simulate the dangers of excessive medication we assume that the medicine 
helps the immune system by increasing the window between $\theta_1$ and
$\theta_2$ within which the immune response is positive. We made a test
where the medication takes effect only after a few interactions (much shorter 
delays than in the aging studies of \cite{bernardes}) and then found
the results to be the same as if the medication effects start with the 
beginning of the simulation. Thus all simulations reported below use one 
time-independent pair of thresholds, with smaller $\theta_2/\theta_1$ 
corresponding to no medication, and larger $\theta_2/\theta_1$ to the effect
of medicine. At what threshold $\theta_2$, for fixed $\theta_1$, does the system
switch from healthy localised to dangerous, huge, spreading immune response?

\section {Results}

Figure 1 shows how the number of $B$ cells per site
varies as a function of the second threshold $\theta_2$ for a  $\theta_1$ 
equal to 10. The antigen presentation was simulated by randomizing one line
of length $L$, while all other sites were set to $B=0$.  The results usually
indicate that for $\theta_2 = 5 \theta_1 $ the $B$
 cell concentration stays small
while for $\theta_2 = 10 \theta_1$ it spreads. This general trend can be 
violated in some cases, see the caption to Fig.1.
The constant value (the plateau) is higher for an
 overdosed drug, meaning that
the system memorises the amount of medicine absorbed. Figure 2 shows two 
phase transitions a a function of $\theta2$.

\begin{figure}[hbt]
\begin{center}
\includegraphics[angle=-90,scale=0.5]{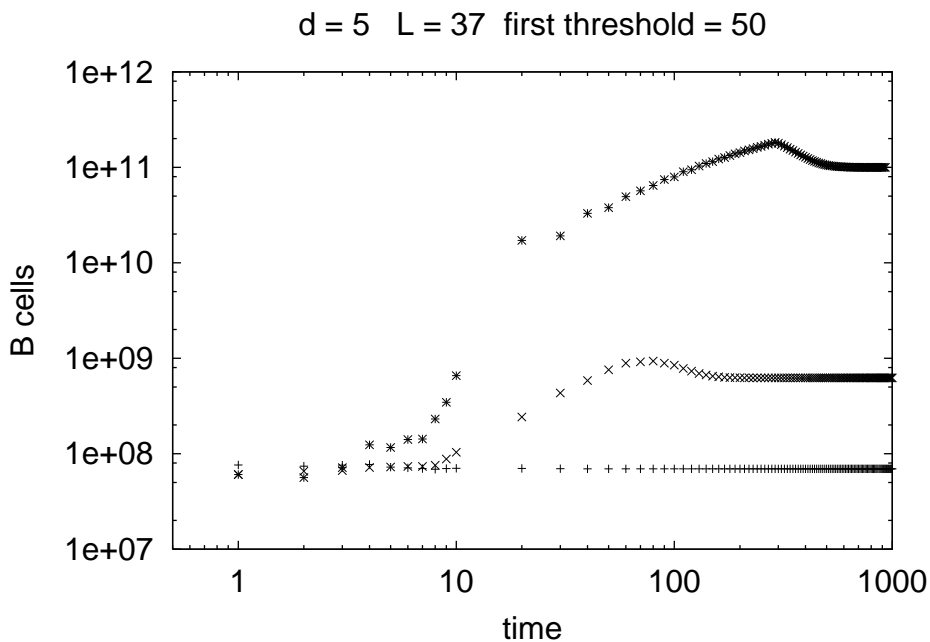}
\end{center}
\caption{Double-logarithmic plot of non-normalised
response $B$ versus
time $t$ for a high value of the first threshold, $\theta_1= 50$.
 A healthy response
of the system is observed  for $R$ = 5, i.e. $\theta_1=250$ (+).
The $R$ = 10 ( $\theta_1=500$) case is shown with (x), and
$R$ = 20 with (*).
}
\end{figure}

\begin{figure}[hbt]
\begin{center}
\includegraphics[angle=-90,scale=0.5]{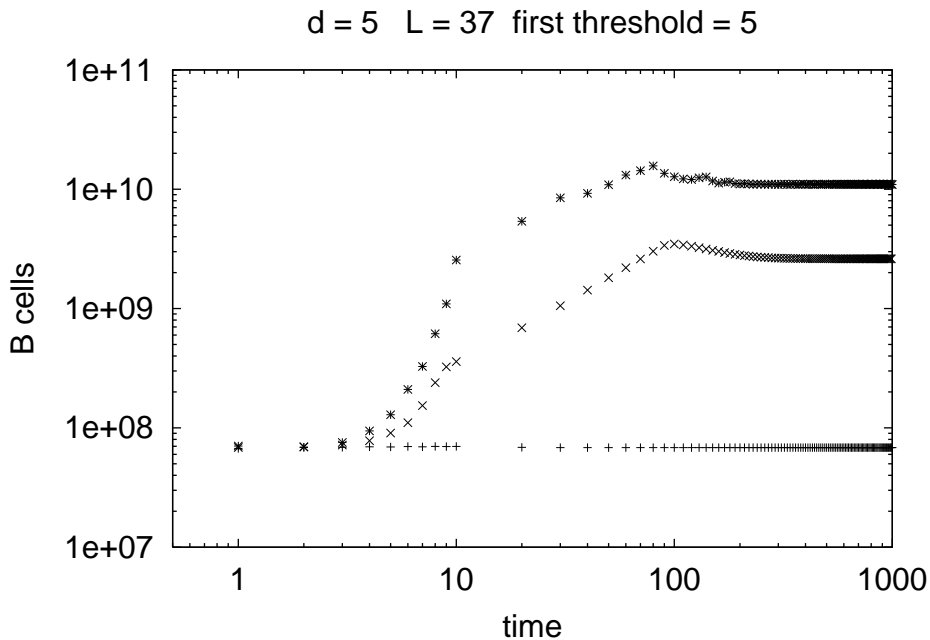}
\end{center}
\caption{Double-logarithmic plot of non-normalised
response $B$ versus
time $t$ for a low value of the first threshold, $\theta_1= 5$.
 A healthy response
of the system is observed  for $R$ = 2, denoted with (+), while $R = 5$
(x) gives a smoothly
increasing curve, similar to the curve for $R = 10$ in Fig.3. The
 $R$=20 case is shown with (*).
}
\end{figure}

\begin{figure}[hbt]
\begin{center}
\includegraphics[angle=-90,scale=0.5]{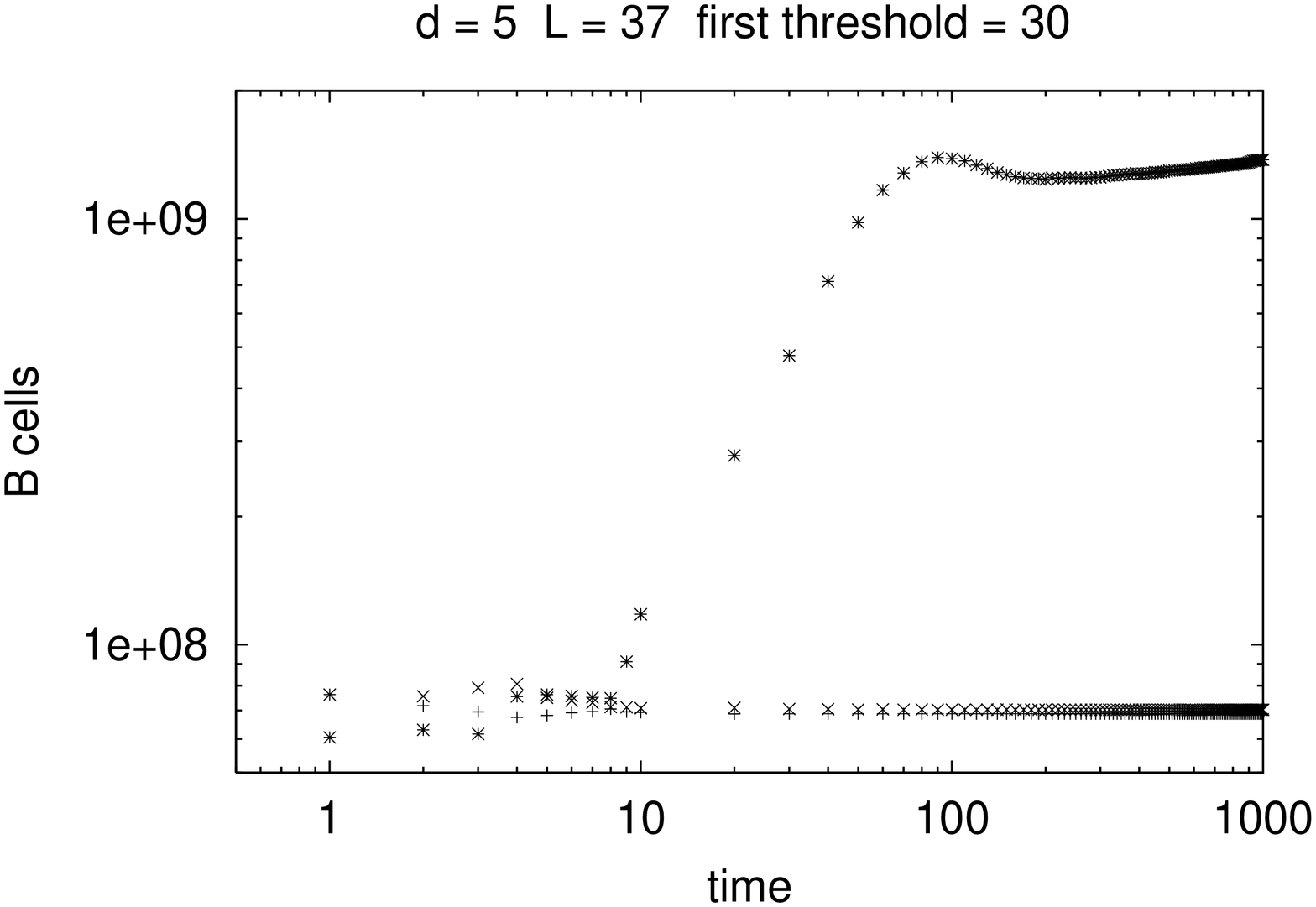}
\end{center}
\caption{Double-logarithmic plot of non-normalised
response $B$ versus
time $t$ for one  'special' case, $\theta_1=30$. The system responses
normally upto $R = 10$ (+,x); for $R = 20$ (*) the plateau is observed
for some time, after then the $B$-cells increase again.
}
\end{figure}

The initial configuration with randomly flipped sites in one hyper-plane
models a higher response than previously.
We observe different time behaviours of the system
depending on 
both the ratio $R=\theta_1/\theta_2$ and the $\theta_1$-value.

If the first threshold $\theta_1$ is 50, then the number of $B$-cells
remains basically unchanged for $R$ less or equal to 5, 
increases smoothly for $R$ about 10, or has a prominent peak for $R$ about
20. The time at which the number of $B$ cells reaches its maximum  value
increases with the $\theta_2$-value, Fig.3.

A similar study with  small $\theta_1$-values (5,10) shows an
opposite time dependence: 
the peak appears at shorter times with the $R$-increase, Fig.4.
However, the shapes change  similarly, i.e. a smooth increase to the
maximum $B$-value for $R = 5, 10$. Note that a healthy response (one
$B$ cell per a site) is observed 
for much smaller $R$-values, about 2, e.g. for a narrower window.
This is a general trend, the smaller $\theta_1$-value, the narrower window
for a healthy response.

A different time-dependence of the immune response is sometimes observed:
for example, if $\theta_1$ = 30 and $R$ = 20,
 the number of the $B$ cells increases
again after some time being almost constant (at a plateau), Fig.5. Thus it seems
that the model is able to describe the 'unhealthy' response of the immune system
to drugs if a specific time of treatment is surpassed.

\section{Summary}
Using a variant of the well-established BSP model we show that a properly
working immune system may go wrong completely and attack ``everything'' if
the second threshold is increased too much. This phase transition may
explain some autoimmune diseases arising from medication.

\bigskip 
This work was supported by the Sofia-Cologne university partnership.
We thank R.M. Zorzenon dos Santos for encouragement.

\end{document}